\documentclass[twocolumn,pra,aps,amssymb]{revtex4}

\usepackage{graphicx}
\usepackage{epstopdf}
\usepackage{todonotes}

\usepackage{amsmath,amssymb,amsfonts,amsthm}
\usepackage{multirow}
\usepackage{verbatim}
\usepackage{url}
\usepackage{comment}
\usepackage{mathtools}
\usepackage{epsf,pgf,graphicx}
\usepackage{color}
\usepackage{tikz,pgf}
\usepackage{makecell}
\usepackage{color}
\usepackage{pdfpages}
\newtheorem{lemma}{Lemma}

\newcommand{\ket}[1]{\ensuremath{\left|#1\right\rangle}}

\newcommand{\sio}{SIMON }

\usepackage[utf8]{inputenc}

\DeclareFixedFont{\ttb}{T1}{txtt}{bx}{n}{9} 
\DeclareFixedFont{\ttm}{T1}{txtt}{m}{n}{9}  

\usepackage{color}
\definecolor{deepblue}{rgb}{0,0,0.5}
\definecolor{deepred}{rgb}{0.6,0,0}
\definecolor{deepgreen}{rgb}{0,0.5,0}

\usepackage{listings}

\newcommand\pythonstyle{\lstset{
		language=Python,
		basicstyle=\ttm,
		morekeywords={self},              
		keywordstyle=\ttb\color{deepblue},
		emph={MyClass,__init__},          
		emphstyle=\ttb\color{deepred},    
		stringstyle=\color{deepgreen},
		frame=tb,                         
		showstringspaces=false
}}

\lstnewenvironment{python}[1][]
{
	\pythonstyle
	\lstset{#1}
}
{}

\newcommand\pythoninline[1]{{\pythonstyle\lstinline!#1!}}

\begin{document}
\title{Linear Cryptanalysis through the Lens of Clauser-Horne-Shimony-Holt Game}
\author{Arpita Maitra$^1$, Ravi Anand$^2$, Suman Dutta$^3$} 
\affiliation{$^1$TCG Centre for Research and Education in Science and Technology, \\ Kolkata-700091, West Bengal, India, 
$^2$R C Bose Centre for Cryptology and Security, Indian Statistical Institute, 203 B.T. Road, Kolkata 700108, West Bengal, India,
$^3$ Applied Statistic Unit, Indian Statistical Institute, 203 B.T. Road, Kolkata-700108, West Bengal, India}
\email{arpita76b@gmail.com\\ ravianandsps@gmail.com\\ sumand.iiserb@gmail.com}
\begin{abstract}
Application of CHSH game in Linear Cryptanalysis is presented. Till date, the known usage of CHSH game in Quantum Cryptology is to verify the device independence of the protocols. We observed that the game can be exploited to improve the bias of some ciphers and hence can be used in cryptanalysis. In the present initiative, we showed the application of the game in linear cryptanalysis on a lightweight cipher named {\em SIMON}. This observation opens a new direction of research in quantum cryptography. 
\end{abstract}
\maketitle 

\section{Introduction} 
Bell inequality~\cite{bell} certifies the non-local relationship between two correlated systems. In other words, Bell inequality confirms the quantumness of the system. This inequality can be described in form of a game, named CHSH~\cite{chsh} game.

In CHSH game, there are one referee and two players whom we generally call Alice and Bob. Referee provides a bit $x$ to Alice and a bit $y$ to Bob. After receiving the bits, Alice and Bob each outputs a bit, say, $a$ for Alice and $b$ for Bob. The players win the game if $x\wedge y=a\oplus b$.  

In classical domain where there is no existence of entanglement, the best strategy to win the game would be as follows.
\begin{enumerate}
\item Alice and Bob both output $0$ irrespective of their inputs.
\item Alice and Bob both output $1$ irrespective of their inputs.
\end{enumerate}
In such case, they will win the game with probability $0.75$. 
On the other hand, in quantum domain where one can take the advantage of entanglement, wins the game with better probability.

 In this case, Alice and Bob share maximally entangled state prior to the game. After receiving the inputs from referee, the players measure their respective systems in some specified bases. The basis choice depends on the input bits. It is well proven that in such strategy, Alice and Bob can improve their winning probability upto $0.85$.  

In quantum cryptography, CHSH game is used for testing the device independence of the protocols. Precisely, $0.85$ probability certifies the existence of maximally entangled state amongst the legitimate parties~\cite{Mayer,acin06a,acin06b,scarani06,VV,lim,giustina}. If the shared states are maximally entangled, then from the monogamy relation of entanglement~\cite{tomamichel,ruv13,terhal}, it is guaranteed that the information about the raw key extracted by an eavesdropper, generally familiar as Eve, can not be greater than the information extracted by Bob (one of the authenticated parties). 

The probability is calculated from input-output statistics. For example, in case of Quantum Key Distribution (QKD) protocols, the QKD boxes are available as Black boxes to the legitimate parties. Each box can take an input bit ($x$ or $y$) and provides an output bit ($a$ or $b$). If the input-output statistics satisfies the winning condition with probability $0.85$, then there must exists a non-local maximal correlation between the boxes available to the authenticated parties. This non-local correlation guarantees the absolute security of the protocol. 
In the present initiative, we report another application of the game in quantum cryptography. 

In cryptography, cipher is the encrypted form of message.  Message is encrypted in the motivation towards hiding the information from the eavesdropper. The encryption as well as the decryption function are known to all, however the secret thing is the key. The encryption function is chosen in such a way so that without any knowledge of the key it is not possible to decrypt the cipher. Even the designer of the crypto-system also can not decipher it without the knowledge of the key. Hence, extracting the key bits is the prime motivation for attacking the cipher. Formally, we name it as {\em cryptanalysis} (breaking the code).

It is expected that the ciphers are designed in such a way so that the random variables $0$ and $1$ come from a probability distribution $\{1/2,1/2\}$. However, any shifting from this distribution may cause loopholes in the cipher. Exploiting this one can extract the key bit(s). The deviation from $1/2$ is called bias. Thus, finding a bias for a given cipher is another important job in cryptanalysis. 

Now, consider that there are two probability distributions $\{p, 1-p\}$ and $\{q,  1-q\}$. To distinguish these two probability distributions with confidence very close to $1$, the number of samples required is approximately $1/pq^2$. If the value of $q$ increases, then the number of samples decreases. If $p=1/2$ and $q=1/2+\epsilon$, then $\epsilon$ is called bias for the random variables. Hence, improving the bias is very significant contribution in cryptanalysis. 

In the present draft, we observe that if we can convert the classical ciphers in quantum ciphers, i.e, bits will be converted to qubits and the classical gates used in the ciphers will be converted to the quantum gates, then we may exploit quantum advantage to improve the bias present in the classical cipher. Here, we consider a lightweight cipher named {\em SIMON}~\cite{simon}. It is a symmetric cipher, i.e, the sender and the receiver use the same secret key. \sio is a lightweigh cipher designed by NSA (National Security Agency of USA). That is why this cipher is very important in the domain of classical cryptography. 

In classical domain, it was observed that this cipher has some bias and the optimal bias found is $0.25$ i.e., here $q=0.75$. On the other hand, exploiting Boolean version of CHSH game, we can successfully improve the bias upto $0.35$, i.e., in this case, $q$ will be $0.85$.

Till date, Quantum Cryptanalysis on Symmetric Ciphers exploits Grover's search algorithm~\cite{grover}, Simon period finding algorithm~\cite{Simon1} and the combination of both.  Block ciphers like AES~\cite{aesg, aes1, aes2, aes3}, Even-Mansour construction~\cite{fx} and McEliece system~\cite{eliece} have been studied subsequently. It is also proven that classically secure ciphers can be broken with quantum algorithms~\cite{3roun, kaplan, kaplan1,demeric,fx}. Quantum algorithm can be used to speed up classical attacks~\cite{kaplan2,hs,tho} too.

Contrary to these, in the current manuscript, we exploit CHSH game and improve the bias of the cipher.  This improvement should have major impact in linear and differential cryptanalysis~\cite{stinson}. We believe that this will open up a new avenue of research in the paradigm of quantum cryptography .

The draft is organized as follows. Section II deals with a brief description of classical cipher {\em SIMON}, its encryption and decryption algorithms, existing cryptanalysis on the cipher. In Section III, we discuss Boolean implementation of CHSH game. In section Section IV we show how the Boolean circuit of the game can be exploited in {\em SIMON}. We implement the idea in IBMQ simulator for an arbitrary round $i$. Section V concludes the paper.

\section{Brief description of \sio}
\sio is a family of balanced Feistel structured~\cite{stinson} lightweight block ciphers with 10 different block sizes and key sizes (Table~\ref{Spara}). 
\begin{table}[ht]
		\centering
		\scalebox{0.75}{
			\begin{tabular}{|c|c|c|c|c|}
				\hline
				Block Size $(2n)$ & Key Size $(k=mn)$ & word size $(n)$  & keywords $(m)$ & Rounds $(T)$ \\
				\hline
				32 & 64 & 16 & 4 & 32\\
				\hline
				48 & 72,96 & 24 & 3,4 & 36,36\\
				\hline
				64 & 96,128 & 32 & 3,4 & 42,44\\
				\hline
				96 & 96, 144 & 48 & 2,3 & 52,54\\
				\hline
				128 & 128,192,256 & 64 & 2,3,4 & 68,69,72\\
				\hline
		\end{tabular}}
		\caption{\sio parameters}
		\label{Spara}
	\end{table}

\begin{figure}[ht]
	\centering
	\includegraphics[scale = 0.6]{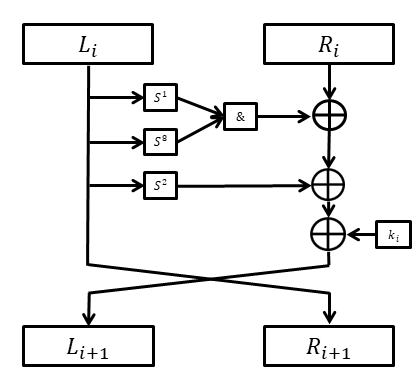}
	\caption{\sio round function}
	\label{sim}
\end{figure}

The round function used in the Feistel structure of \sio block ciphers consists of circular shift, bitwise AND and bitwise XOR operations. The state update function is defined as,
\begin{align}
F(x,y) = (y \oplus S^1 (x) S^8 (x) \oplus S^2 (x) \oplus k, x)
\label{supdate}
\end{align}
The structure of one round \sio encryption is depicted in Figure~\ref{sim}, where $ S^j $ represents a left circular shift by $ j$ bits, $L_i$ and $ R_i $ are $n$-bit words which constitutes the state of \sio at the $i$-th round and $ k_i $ is the round key which is generated by key scheduling algorithm. The description is out of the scope for the paper. Interested readers may explore~\cite{simon} for detailed description of the lightweight ciphers, \sio and SPECK.\\ 

	Let $(L_0,R_0)$ be the initial state and the state propagate as $(L_0,R_0),(L_1,R_1),(L_2,R_2), \cdots , (L_T,R_T)$ upto $T$ rounds. If $K$ is the key for a round $i$,  then from the state update function of \sio, one can write the following.
	\begin{eqnarray*}
	R_{i+2}(j)&=& L_{i+1}(j)\\
	&=& R_i(j)\oplus K_i(j)\oplus (L_i(j+1) \& L_i(j+8))\oplus L_i(j+2), \\
	\end{eqnarray*}
	where, $0\leq i \leq (n/2)$.
	Here $i$ stands for round number and $j \mod (n/2)$ denotes the position of the bit. 
	
	Now, we consider $L_i(j)\oplus L_{i+1}(j)$. That is we are establishing a linear relationship between the $j$th bit of $i$th round and the $j$th bit of $i+1$ round. In other words, we are establishing a linear relation between a bit and its encrypted cipher bit. Due to the encryption function (round update function) of \sio , this will be equal to the following expression.
	\begin{eqnarray*}
	L_i(j)\oplus R_i(j)\oplus K_i(j)\oplus (L_i((j+1)) \& L_i((j+8)))\\
	 \oplus L_i(j+2))
	\end{eqnarray*}	
	
The Boolean function `AND' (\&) plays a vital role in this linear relationship. A close observation reveals that if $L_i(j)=L_i(j+2)$, then due to the presence of `AND' operation we can write the followings.
\begin{align*}
\Pr(L_i(j)\oplus L_{i+1}(j)=0|R_i(j)=K_i(j))=3/4\\
\Pr(L_i(j)\oplus L_{i+1}(j)=1|R_i(j)=K_i(j))=1/4
\end{align*}
These imply that there is certain bias in the cipher. Hence, for a sufficient number of plaintext-ciphertext pairs for a round $i$, it is possible to find such bias. This can be further extended for two conjecutive rounds, i.e., for $i$ and $i+2$. This is because of the advantage of Feistel construction. In Feistel construction, $L_{i+1}(j)=R_{i+2}(j)$. Hence, it is rather better to say that for a sufficient number of plaintext-ciphertext pairs for two rounds {\em SIMON}, it is possible to find the above bias. 

The procedure would be the following.
\begin{enumerate}
\item For a given plaintext-ciphertext pairs for two round \sio, check if $L_i(j)=L_{i+1}(j+2)$. Note that $L_{i+1}(j+2)=R_{i+2}(j+2)$.
\item Calculate $\Pr(L_i(j)\oplus L_{i+1}(j)=0)$ and $\Pr(L_i(j)\oplus L_{i+1}(j)=1)$.
\item Iff $\Pr(L_i(j)\oplus L_{i+1}(j)=0)=3/4$ or $\Pr(L_i(j)\oplus L_{i+1}(j)=1)=1/4$, conclude $K_i(j)=R_i(j)$.
\end{enumerate}
The variants of such attack has been studied in~\cite{somitra,Abed,Alkhzaimi}. However, the bias better than $0.25$ for a round has not been found yet. In this backdrop, we observed that as the bias comes from `AND' operation, one can take the advantage of CHSH game and hence can improve the bias for a round. Before going to the implementation of CHSH game in \sio, we like to discuss the Boolean implementation of CHSH game which is the pilar of improving the bias.
\section{Boolean implementation of CHSH game}
In this section, CHSH game will be viewed as a Boolean function $f:\{0,1\}^n\rightarrow \{0,1\}$. Here, the function $f:\{0,1\}^4\rightarrow\{0,1\}$ and $f=(x\wedge y)\oplus (a\oplus b)$. The corresponding circuit is given below.
\begin{figure}[ht]
	\centering
	\includegraphics[scale = 0.4]{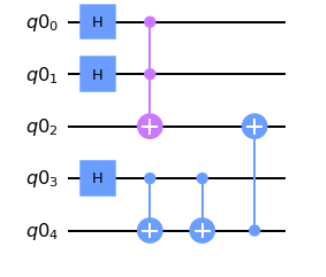}
	\caption{Boolean version of classical CHSH game}
	\label{chsh}
\end{figure}
In the circuit, $q0_0=x$, $q0_1=y$, $q0_3=a$ and $q0_4=b$. Qubit $q0_2$ stores the functional value $f$. 

According to the classical strategy for winning the game with maximum probability, Alice and Bob output the same bits. This has been depicted using first CNOT gate (from left) as a cloning operator on $q0_3$. It copies $q0_3$ in $q0_4$ resulting $a=b$ always. Hadamard gates are used to take care of all possible choices of $x$, $y$ and $a$. If we compute the probability considering the reduced density matrix of $q0_2$, we will get $0.75$ for the bit 0 and $0.25$ for the bit 1.

\begin{figure}[ht]
	\centering
	\includegraphics[scale = 0.4]{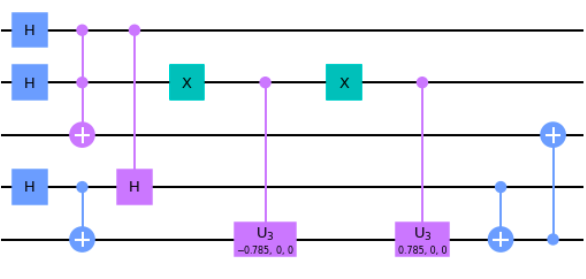}
	\caption{Boolean version of quantum CHSH game}
	\label{chsh1}
\end{figure}

Now, to implement quantum strategy we took the help of Controlled-Hadamard ($CH$) and Controlled-U3 ($CU_3$) gate, where $U_3=
\begin{pmatrix}
\cos{\theta/2} & -\sin{\theta/2}\\
\sin{\theta/2} & \cos{\theta/2}
\end{pmatrix}
$. The corresponding circuit is given in Fig~\ref{chsh1}.
According to the quantum strategy, Alice and Bob share maximally entangle state prior to the game. Applying Hadamard gate on $q0_3$ followed by CNOT on both $q0_3$ and $q0_4$, where $q0_3$ is the controlled bit, we generate the maximally entangle state $\frac{1}{\sqrt{2}}(\ket{00}+\ket{11})$ between Alice and Bob. Controlled-Hadamard serves Alices's strategy, i.e., if $x=0$, Alice will measure her sub-system in $\{\ket{0},\ket{1}\}$ basis (no Hadamard applied ), else she will measure it in $\{\ket{+},\ket{-}\}$ basis (Hadamard applied), where $\ket{+}=\frac{1}{\sqrt{2}}(\ket{0}+\ket{1})$ and $\ket{-}=\frac{1}{\sqrt{2}}(\ket{0}-\ket{1})$. Similarly, $CU_3$ serves Bob's strategy, i.e., if $y=0$, Bob will measure his sub-system in $\{\ket{\pi/8},\ket{-\pi/8}\}$ basis (first $CU_3$ from left) and else in  $\{\ket{3\pi/8},\ket{-3\pi/8}\}$ basis (second $CU_3$ from left), where $\ket{\pi/8}=\cos(\pi/8)\ket{0}+\sin(\pi/8)\ket{1}$, $\ket{-\pi/8}=-\sin(\pi/8)\ket{0}+\cos(\pi/8)\ket{1}$, and $\ket{3\pi/8}=\cos(3\pi/8)\ket{0}+\sin(3\pi/8)\ket{1}$, $\ket{-3\pi/8}=-\sin(3\pi/8)\ket{0}+\cos(3\pi/8)\ket{1}$. Note that $\pi/4=0.785$. That is why we set $\theta$ as $0.785$ when $U_3$ has been constructed. We compute the probability considering the reduced density matrix of $q0_2$. We obtained $0.85$ (taking two decimal places) for the bit 0 and $0.15$ for the bit 1.

In the following section, we will show how we exploit this Boolean circuit in \sio and improve the bias. 
\section{Boolean CHSH game for improving the bias of \sio}
\begin{figure}[ht]
	\centering
	\includegraphics[scale = 0.4]{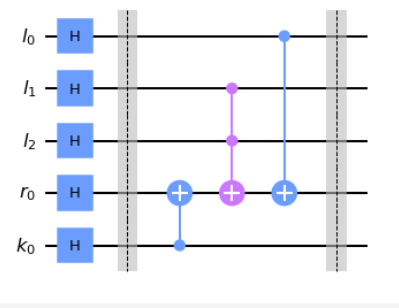}
	\caption{Quantum version of \sio round update function}
	\label{chsh2}
\end{figure}

\begin{figure}[ht]
	\centering
	\includegraphics[scale = 0.4]{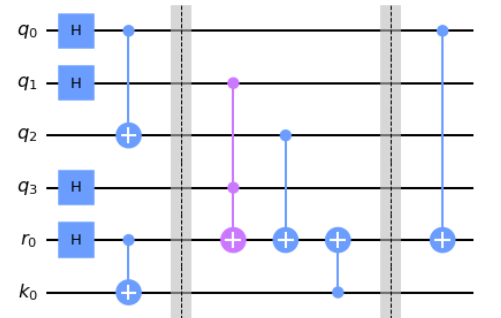}
	\caption{Quantum version of linear approximation} 
	\label{chsh3}
\end{figure}

To exploit the Boolean circuit of CHSH game in \sio, we need to design \sio round function in quantum domain. 	
We assume that we have $k$-qubits reserved for the key, $K$, and $n$-qubits each for $L$ and $R$. The classical Boolean operations are now replaced by the quantum reversible gates. The replacement is as follows. For detail explanation one may explore~\cite{qinp}.
\begin{enumerate}
		\item `AND' is replaced by {\em Toffoli}
		\item `XOR' is replaced by CNOT
	\end{enumerate}
	Thus the quantum version of \sio round function will be as follows.
	\begin{eqnarray*}
	F(x,y) &=& CNOT((CNOT(K_i(j),R_i(j)), \\
	&&CNOT((Toffoli (L_i(j+1), \\
	&&L_i(j+8), R_i(j)), L_i(j+2))))
	\end{eqnarray*}
where, $L_i(j+m)$ represents $m$ bits shift on $L_i(j)$, i.e., $S^{m}(L_i(j))$. The corresponding circuit is presented in Fig~\ref{chsh2}.	
Here, $l_0$, $l_1$ and $l_2$ represent $L_i(j+2)$, $L_i(j+1)$ and $L_i(j+8)$ respectively. $r_0$ represents $R_i(j)$ and $k_0$ stand for $K_i(j)$. Hadamard gates are taking care of all possible input states. In this case, there is no bias found, i.e, probability of the random variable 0 is $1/2$ and probability of the random variable 1 is also $1/2$. However, 
we observe that if for a round $i$, $L_i(j)=L_i(j+2)$, then the followings happen. Such situation is picturized by the circuit given in Fig~\ref{chsh3}.
\begin{align*}
\Pr(L_i(j)\oplus L_{i+1}(j)=0|R_i(j)=K_i(j))=3/4\\
\Pr(L_i(j)\oplus L_{i+1}(j)=1|R_i(j)=K_i(j))=1/4
\end{align*}
In this circuit, $q_0$, $q_1$, $q_2$ and $q_3$ represent $L_i(j)$, $L_i(j+1)$, $L_i(j+2)$ and $L_(j+8)$ respectively. As $L_i(j)=L_i(j+2)$, we put CNOT gate on $q_0$ and $q_2$ considering $q_0$ as control bit. CNOT gate operates as a cloning machine here. $q_0$ is copied into $q_2$ making $L_i(j)=L_i(j+2)$. Similarly, as $R_i(j)=K_i(j)$, we again apply CNOT gate on $r_0$ so that it will be copied into $k_0$. Other parts of the circuit remain same. After the second barrier, we apply CNOT gate on $q_0$ and $r_0$ to take care of the linear relation between $L_i(j)$ and $L_{i+1}(j)$, i.e., $L_i(j)\oplus L_{i+1}(j)$.  

\begin{figure}[ht]
	\centering
	\includegraphics[scale = 0.3]{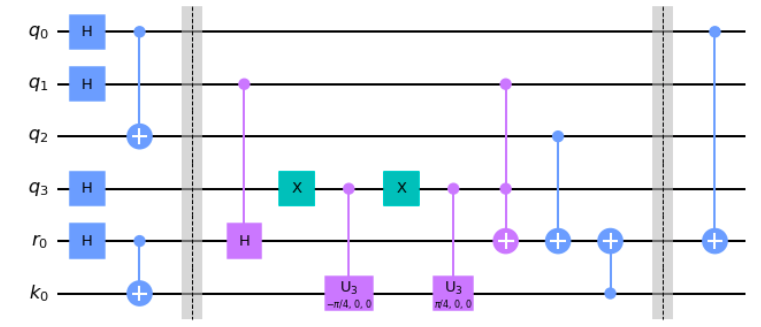}
	\caption{Modified version of linear approximation}
	\label{chsh4}
\end{figure}

We modify the circuit with Boolean CHSH game and observe that $\Pr(L_i(j)\oplus L_{i+1}(j)=0|R_i(j)=K_i(j))=0.85$ and $\Pr(L_i(j)\oplus L_{i+1}(j)=1|R_i(j)=K_i(j))=0.15$. The modified circuit is given in Fig~\ref{chsh4}. We also prove that the optimal probability is $0.85$. Calculations for different choices of $q_1$ and $q_3$, i.e., for all possible options between $L_i(j+1)$ and $L_i(j+8)$ are shown in the Appendix. We do not consider other bits, as in this case, $L_i(j)=L_i(j+2)$ and $K_i(j)=R_i(j)$.
We have implemented all the circuits in IBMQ interface. For the experiments, we use IBMQ simulator. The codes for the linear approximation and its modified version are also provided in~\ref{app}. 

\section{Discussions and Conclusion} In the present initiative, we have shown how to improve the bias of a classical lightweight cipher called \sio in linear cryptanalysis exploiting CHSH game. In the domain of {\em Quantum Cryptanalysis on Symmetric Ciphers} we mainly study the impact of Grover's search algorithm~\cite{grover}, Simon period finding algorithm~\cite{Simon1} and the combination of both on block ciphers. It is also shown that classically secure ciphers can be broken with quantum algorithms.
Quantum algorithm can be used to speed up classical attacks too. 

Contrary to these, in the present draft, we present different kind of attack where bias of a random variable has been improved. This should have major impact in linear and differential cryptanalysis. 

Here, we consider lightweight cipher \sio only. However, we can construct a lot many CHSH like games which can be used to improve the bias of classical symmetric ciphers: stream as well as block ciphers. The present finding is the tip of the iceberg. This finding will open-up new direction of research in the domain of quantum cryptography. 

Finally, the result again proves the quantum advantage over classical domain showing that in quantum regime, the symmetric ciphers are not as secure as those are in classical domain. We may need less samples to distinguish two probability distribution functions $p$ and $q$. This signals that we should be very careful when designing a cipher for post quantum era. 

Our future research will be directed towards linear and differential cryptanalysis of different symmetric ciphers using the result presented in the paper as well as designing various CHSH like games which might be exploited to improve the bias of existing ciphers.

\section{Appendix} 
\label{app}
We begin this section by providing the mathematical derivations for modified linear approximations using CHSH game.
We look into the evolution of qubits $\ket{r_0k_0}$ corresponding to the quantum circuit given in Figure 6. For a better readability, the same quantum circuit is provided again. 
\begin{figure}[ht]
	\centering
	\includegraphics[scale = 0.33]{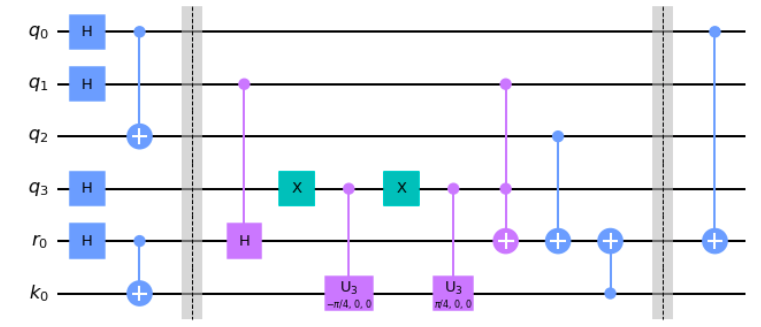}
\end{figure}
Observe that, before the first barrier, $\ket{r_0k_0}$ is in a maximally entangled state, $\frac{1}{\sqrt{2}}\left(\ket{00}+\ket{11} \right)$. Here we consider four different possibilities for $q_1$ and $q_3$, namely, $q_1q_3=00,01,10$ or $11$ and measure the modified bias from the qubit $\ket{r_0}$. Notice that, since $q_0=q_2$ and both $q_2$ (before the second barrier) and $q_0$ (at the end ) are being XOR-ed with $r_0$ without any modification, they will nullify each other and we can skip them while studying the state evolution.\\
\\
\textit{\textbf{Case I.}} $q_1=0, \, q_3=0$:\\
Since $q_1=q_3=0$, the $CH$ gate controlled by $q_1$ and the $CCX$ gate controlled by both $q_1,q_3$ will not be activated here. Thus the starting state evolves as follows.
\begin{widetext}
\begin{align*}
\ket{r_0k_0}=\frac{1}{\sqrt{2}}\left(\ket{00}+\ket{11} \right)
\xrightarrow{I\otimes U\left(-\theta,0,0 \right)}&\frac{1}{\sqrt{2}}\left[\ket{0}\left(\cos{\frac{\theta}{2}}\ket{0}- \sin{\frac{\theta}{2}}\ket{1}\right)+\ket{1}\left(\sin{\frac{\theta}{2}}\ket{0} + \cos{\frac{\theta}{2}}\ket{1}\right) \right]\\
=&\frac{1}{\sqrt{2}}\left( \cos{\frac{\theta}{2}}\ket{00}-\sin{\frac{\theta}{2}}\ket{01}+\sin{\frac{\theta}{2}}\ket{10}+\cos{\frac{\theta}{2}}\ket{11}\right)\\
\xrightarrow{CX_{k_0,r_0}}&\frac{1}{\sqrt{2}}\left( \cos{\frac{\theta}{2}}\ket{00}-\sin{\frac{\theta}{2}}\ket{11}+\sin{\frac{\theta}{2}}\ket{10}+\cos{\frac{\theta}{2}}\ket{01}\right).
\end{align*}

Therefore the probability is given by, $\displaystyle\Pr\left(r_0=0\right)=\frac{1}{2}\left(\cos^2{\frac{\theta}{2}}+\cos^2{\frac{\theta}{2}} \right)=\cos^2{\frac{\theta}{2}}.$\\
\end{widetext}
\textit{\textbf{Case II.}} $q_1=0, \, q_3=1$:\\
Since $q_1=0$, the $CH$ gate controlled by $q_1$ and the $CCX$ gate controlled by both $q_1,q_3$ will not be activated here. Therefore, the starting state $\ket{r_0k_0}$ evolves as follows.
\begin{widetext}
\begin{align*}
\ket{r_0k_0}=\frac{1}{\sqrt{2}}\left(\ket{00}+\ket{11} \right)
\xrightarrow{I\otimes U\left(\theta,0,0 \right)}&\frac{1}{\sqrt{2}}\left[\ket{0}\left(\cos{\frac{\theta}{2}}\ket{0}+ \sin{\frac{\theta}{2}}\ket{1}\right)+\ket{1}\left(-\sin{\frac{\theta}{2}}\ket{0} + \cos{\frac{\theta}{2}}\ket{1}\right) \right]\\
=&\frac{1}{\sqrt{2}}\left( \cos{\frac{\theta}{2}}\ket{00}+\sin{\frac{\theta}{2}}\ket{01}-\sin{\frac{\theta}{2}}\ket{10}+\cos{\frac{\theta}{2}}\ket{11}\right)\\
\xrightarrow{CX_{k_0,r_0}}&\frac{1}{\sqrt{2}}\left( \cos{\frac{\theta}{2}}\ket{10}+\sin{\frac{\theta}{2}}\ket{01}-\sin{\frac{\theta}{2}}\ket{00}+\cos{\frac{\theta}{2}}\ket{11}\right).
\end{align*}

Thus the probability is given by
$\displaystyle\Pr\left(r_0=0\right)=\frac{1}{2}\left(\cos^2{\frac{\theta}{2}}+\cos^2{\frac{\theta}{2}} \right)=\cos^2{\frac{\theta}{2}}.$\\
\end{widetext}
\textit{\textbf{Case III.}} $q_1=1, \, q_3=0$:\\
Here the $CH$ gate controlled by $q_1$ will be activated. However, since $q_3=0$ the $CCX$ gate controlled by both $q_1,q_3$ will not be activated here. Hence the state evolution will be as follows.
\begin{widetext}
\begin{align*}
\ket{r_0k_0}=&\frac{1}{\sqrt{2}}\left(\ket{00}+\ket{11} \right)\\
\xrightarrow{H\otimes I}&\frac{1}{2}\left(\ket{00}+\ket{01}+\ket{10}-\ket{11} \right)\\
\xrightarrow{I\otimes U\left(-\theta,0,0 \right)}&\frac{1}{2}\left[\ket{0}\left(\cos{\frac{\theta}{2}}\ket{0}- \sin{\frac{\theta}{2}}\ket{1}\right)+\ket{0}\left(\sin{\frac{\theta}{2}}\ket{0} + \cos{\frac{\theta}{2}}\ket{1}\right)\right.\\
&\hspace{5cm}\left.+\ket{1}\left(\cos{\frac{\theta}{2}}\ket{0}- \sin{\frac{\theta}{2}}\ket{1}\right) - \ket{1}\left(\sin{\frac{\theta}{2}}\ket{0} + \cos{\frac{\theta}{2}}\ket{1}\right)\right]\\
=&\frac{1}{2}\left[ \left(\cos{\frac{\theta}{2}}+\sin{\frac{\theta}{2}}\right)\ket{00}+\left(\cos{\frac{\theta}{2}}-\sin{\frac{\theta}{2}}\right)\ket{01}+\left(\cos{\frac{\theta}{2}}-\sin{\frac{\theta}{2}}\right)\ket{10}-\left(\cos{\frac{\theta}{2}}+\sin{\frac{\theta}{2}}\right)\ket{11}\right]\\
\xrightarrow{CX_{k_0,r_0}}&\frac{1}{2}\left[ \left(\cos{\frac{\theta}{2}}+\sin{\frac{\theta}{2}}\right)\ket{00}+\left(\cos{\frac{\theta}{2}}-\sin{\frac{\theta}{2}}\right)\ket{11}+\left(\cos{\frac{\theta}{2}}-\sin{\frac{\theta}{2}}\right)\ket{10}-\left(\cos{\frac{\theta}{2}}+\sin{\frac{\theta}{2}}\right)\ket{01}\right]
\end{align*}

Therefore,
$\displaystyle\Pr\left(r_0=0\right)=\frac{1}{4}\left[\left(\cos{\frac{\theta}{2}}+\sin{\frac{\theta}{2}}\right)^2+\left(\cos{\frac{\theta}{2}}+\sin{\frac{\theta}{2}}\right)^2 \right]=\frac{1}{2}\left(1+\sin{\theta}\right).$\\
\end{widetext}
\textit{\textbf{Case IV.}} $q_1=1, \, q_3=1$:\\
Since both $q_1=q_3=1$, the $CH$ gate controlled by $q_1$ and the $CCX$ gate controlled by $q_1,q_3$, both will be activated and therefore the evolution of the starting state will be as follows.
\begin{widetext}
\begin{align*}
\ket{r_0k_0}=&\frac{1}{\sqrt{2}}\left(\ket{00}+\ket{11} \right)\\
\xrightarrow{H\otimes I}&\frac{1}{2}\left(\ket{00}+\ket{01}+\ket{10}-\ket{11} \right)\\
\xrightarrow{I\otimes U\left(\theta,0,0 \right)}&\frac{1}{2}\left[\ket{0}\left(\cos{\frac{\theta}{2}}\ket{0}+ \sin{\frac{\theta}{2}}\ket{1}\right)+\ket{0}\left(-\sin{\frac{\theta}{2}}\ket{0} + \cos{\frac{\theta}{2}}\ket{1}\right)\right.+\\
&\hspace{5cm}\left.\ket{1}\left(\cos{\frac{\theta}{2}}\ket{0}+ \sin{\frac{\theta}{2}}\ket{1}\right) - \ket{1}\left(-\sin{\frac{\theta}{2}}\ket{0} + \cos{\frac{\theta}{2}}\ket{1}\right)\right]\\
=&\frac{1}{2}\left[ \left(\cos{\frac{\theta}{2}}-\sin{\frac{\theta}{2}}\right)\ket{00}+\left(\cos{\frac{\theta}{2}}+\sin{\frac{\theta}{2}}\right)\ket{01}+\left(\cos{\frac{\theta}{2}}+\sin{\frac{\theta}{2}}\right)\ket{10}-\left(\sin{\frac{\theta}{2}}-\cos{\frac{\theta}{2}}\right)\ket{11}\right]\\
\xrightarrow[q_1=q_3=1]{CCX_{q_1,q_3,r_0}}&\frac{1}{2}\left[ \left(\cos{\frac{\theta}{2}}-\sin{\frac{\theta}{2}}\right)\ket{10}+\left(\cos{\frac{\theta}{2}}+\sin{\frac{\theta}{2}}\right)\ket{11}+\left(\cos{\frac{\theta}{2}}+\sin{\frac{\theta}{2}}\right)\ket{00}-\left(\sin{\frac{\theta}{2}}-\cos{\frac{\theta}{2}}\right)\ket{01}\right]\\
\xrightarrow{CX_{k_0,r_0}}&\frac{1}{2}\left[ \left(\cos{\frac{\theta}{2}}-\sin{\frac{\theta}{2}}\right)\ket{10}+\left(\cos{\frac{\theta}{2}}+\sin{\frac{\theta}{2}}\right)\ket{01}+\left(\cos{\frac{\theta}{2}}+\sin{\frac{\theta}{2}}\right)\ket{00}-\left(\sin{\frac{\theta}{2}}-\cos{\frac{\theta}{2}}\right)\ket{11}\right]
\end{align*}
Therefore, the probability is given by
$$\Pr\left(r_0=0\right)=\frac{1}{4}\left[\left(\cos{\frac{\theta}{2}}+\sin{\frac{\theta}{2}}\right)^2+\left(\cos{\frac{\theta}{2}}+\sin{\frac{\theta}{2}}\right)^2 \right]=\frac{1}{2}\left(1+\sin{\theta}\right).$$
\end{widetext}

To maximize the probabilities, we can form a function $h(\theta)=\cos^2{\frac{\theta}{2}}+\frac{1}{2}\left(1+\sin{\theta}\right)$ and using the derivative test we can show that $h$ becomes maximum for $\theta=\frac{\pi}{4}$. Therefore, when $\theta=\frac{\pi}{4}$, individual probabilities also becomes maximum and the bias is given by $0.853$.\\

In the next part, we provide the QISKIT codes for implementing the quantum version of SIMON round function. We first design the classical linear approximation, where we obtain the bias to be $0.75$ and then we provide the modified version, proposed by us, where the bias obtained is given by $0.854$. The bias observed from QISKIT simulation is equivalent with the theoretical bias shown above.
The notations we use here are consistent with the main manuscript.

\begin{python}
	%matplotlib inline 
	from qiskit import *
	from qiskit.tools.visualization 
	import plot_histogram, 
	plot_bloch_multivector
	import qiskit.quantum_info as qi
	from qiskit.quantum_info.states 
	import partial_trace
	import numpy as np
	
	# Quantum linear approximation
	q = QuantumRegister(4, 'q')
	r = QuantumRegister(1, 'r')
	k = QuantumRegister(1, 'k')
	qc=QuantumCircuit(q,r,k)
	
	# Initialization
	qc.h(q[0])
	qc.h(q[1])
	qc.cx(q[0],q[2])
	qc.h(q[3])
	qc.h(r)
	qc.cx(r,k)
	qc.barrier()
	
	# Update function
	qc.ccx(q[1],q[3],r)
	qc.cx(q[2],r)
	qc.cx(k,r)
	qc.barrier()
	qc.cx(q[0],r)
	
	# Calculating the Probabilities
	simulator=Aer.get_backend
	    ('statevector_simulator')
	result=execute
	    (qc,backend=simulator).result()
	Statevector=result.get_statevector()
	rho = np.outer
	    (Statevector, np.conj(Statevector))
	trace_system=[0,1,2,3,5]
	rho_sub=qi.DensityMatrix
	    (partial_trace(rho, trace_system)
	       .data)
	probs = rho_sub.probabilities()
	print(probs)
	
	qc.draw(output='mpl')
\end{python}

The above code outputs the respective probabilities as $[0.75,\,0.25]$ and the corresponding quantum circuit is presented in Figure 5.

In the modified version of linear approximation (Figure 6.), the QISKIT code remains same as above except the `Update function' part which can be designed as follows.

\begin{python}
	# Update function
	qc.ch(q[1],r)
	qc.x(q[3])
	qc.cu3(-np.pi/4,0,0,q[3],k)
	qc.x(q[3])
	qc.cu3(np.pi/4,0,0,q[3],k)
	qc.ccx(q[1],q[3],r)
	qc.cx(q[2],r)
	qc.cx(k,r)
	qc.barrier()
	qc.cx(q[0],r)
\end{python}

And finally we obtain the corresponding probabilities to be $[0.85355339,\, 0.14644661]$.\\

Next we give a brief introduction to the Piling-up lemma. For a better understanding readers may refer to the book \textit{Cryptography: Theory and Practice} by \textit{Douglas R. Stinson}.\\

Consider $k$ independent random variables, $X_1,X_2,\ldots , X_k$, taking on values from the set $\{0, 1\}$. Suppose, $\Pr\left(X_i=0 \right)=p_i$ and $\Pr\left(X_i=0 \right)=1-p_i$ denote the individual probability distribution for $X_i$ and $\epsilon_i=p_i-1/2$ denotes the corresponding bias, where $0\leq p_i \leq 1$ for all $i=1,2,\ldots ,k$.
Then, the following result, which gives a formulation for the bias of the random variable
$X_{i_1}\oplus X_{i_2} \ldots \oplus X_{i_k}$ is known as the \textbf{Piling-up lemma}.
\begin{lemma}[Piling-up lemma]
	If $\epsilon_{{i_1}{i_2}\ldots {i_k}}$ denotes the bias of the random variable $X_{i_1}\oplus X_{i_2} \ldots \oplus X_{i_k}$. Then
	$$\epsilon_{{i_1}{i_2}\ldots {i_k}}=2^{k-1}{\displaystyle \prod_{j=1}^{k} \epsilon_{i_j}}$$
\end{lemma}
For the proof of this lemma, readers are requested to refer to the book, mentioned above.
\end{document}